\documentclass[aps,twocolumn,pre,showpacs,groupedaddress,showkeys,amsmath,amssymb,superscriptaddress]{revtex4}
\usepackage{graphicx}
\usepackage{bm}
\usepackage{epic}
\usepackage{eepic}
\usepackage{pifont}
\usepackage[latin1]{inputenc}
\usepackage{rotating}
\usepackage{color}
\usepackage{nicefrac}
\usepackage{ulem}
\input epsf
\input rotate
\usepackage{graphicx}
\begin{document}
\draft
\title{Two-dimensional Vesicle Dynamics under Shear Flow: Effect of Confinement}
\author{Badr Kaoui}
\email{b.kaoui@tue.nl}
\affiliation{Technische Universiteit Eindhoven, Postbus 513, 5600 MB Eindhoven, The Netherlands}
\affiliation{CNRS - Universit\'{e} Joseph Fourier, UMR 5588, Laboratoire Interdisciplinaire de Physique, B.P. 87, F-38402 Saint Martin d'H\`{e}res Cedex, France}
\author{Jens Harting}
\email{j.harting@tue.nl}
\affiliation{Technische Universiteit Eindhoven, Postbus 513, 5600 MB Eindhoven, The Netherlands}
\affiliation{Institut f\"{u}r Computerphysik, Universit\"{a}t Stuttgart, Pfaffenwaldring 27, D-70569 Stuttgart, Germany}
\author{Chaouqi Misbah}
\email{chaouqi.misbah@ujf-grenoble.fr}
\affiliation{CNRS - Universit\'{e} Joseph Fourier, UMR 5588, Laboratoire Interdisciplinaire de Physique, B.P. 87, F-38402 Saint Martin d'H\`{e}res Cedex, France}
\date{\today}
\begin{abstract}
Dynamics of a single vesicle under shear flow between two parallel
plates is studied in two-dimensions using lattice-Boltzmann simulations. We first
present how we adapted the lattice-Boltzmann method to simulate
vesicle dynamics, using an approach known from the immersed boundary
method. The fluid flow is computed on an Eulerian regular fixed mesh
while the location of the vesicle membrane is tracked by a
Lagrangian moving mesh. As benchmarking tests, the known vesicle
equilibrium shapes in a fluid at rest are found and the dynamical
behavior of a vesicle under simple shear flow is being reproduced.
Further, we focus on investigating the effect of the confinement on
the dynamics, a question that has received little attention so far.
In particular, we study how the vesicle steady inclination angle in
the tank-treading regime depends on the degree of confinement. The
influence of the confinement on the effective viscosity of the
composite fluid is also analyzed. At a given reduced volume (the
swelling degree) of a vesicle we find that both the inclination
angle, and the membrane tank-treading velocity decrease with
increasing confinement.  At sufficiently large degree of confinement
the tank-treading velocity exhibits a non-monotonous dependence on
the reduced volume and the effective viscosity shows a nonlinear
behavior.
\end{abstract}
\pacs{47.63.-b, 47.11.-j, 82.70.Uv}
\keywords{
Vesicle, Tank-treading, Confinement, Lattice-Boltzmann Simulations.
}
\maketitle
\section{Introduction}
The study of blood flow at the microscale, i.e. the scale of blood
corpuscules, is an important issue. In recent years this field has
embraced several communities ranging from medical scientists to
mathematicians. Classical continuum approaches of blood flow, dating back
to a century ago at least \cite{Fung}, are based on several assumptions
and approximations that are both difficult to justify or to validate. For
example,  in the microvasculature, where most of the blood flow resistance
takes place, red blood cells (RBCs), which are by far the major component
of blood, have a size which is of the same order as that of the blood
vessel diameter. Thus, one expects that the discrete nature  of blood
should play a decisive  role in microcirculation.  A prominent example is
the Fahraeus--Lindqvist effect: RBCs cross-streamline migration towards
the blood vessel  center  results in a dramatic collapse of blood
viscosity, causing a reduction  of blood flow resistance in the
microvasculature.  Even  in larger blood  vessels (e.g. veins, arteries) a
satisfactory phenomenological  continuum approach is lacking. One may thus
hope that a constitutive law for blood will ultimately emerge from
numerical simulations taking explicitly  into account  the blood elements.
Still blood flow simulation is a challenging task since it requires
solving for the dynamics of both the blood elements and the suspending
fluid (plasma).

Different numerical methods have been developed to study RBCs or
their bimimetic counterparts (represented by vesicles and capsules),
each having its own  advantages and drawbacks. A widely used  method
is the boundary integral method which is based on the use of Green's
function techniques \cite{Pozrikidis}. It has been successfully
applied to vesicles \cite{Kraus1996,Cantat1999,Biros2008,Biben2010}.
The advantage of this method is the high precision. However, except
for special geometries (e.g. unbounded fluid domain, semi-infinite
domain), an appropriate Green's function is not available. This
means that extra integrations over boundaries delimiting the fluid
have to be performed, which increases the computational time
significantly. In addition, this method is valid for Stokes flow
only (no inertia). Other classes of methods are phase-field
\cite{Biben2003,Biben2005,Du2006} and level set approaches
\cite{Maitre2010} which can be applied both in the Stokes and
Navier-Stokes regimes. Their advantage is the ability of handling,
in principle, many particles by just specifying the initial
condition (in any new run with different vesicles number only
specifying initial conditions is required in principle!). However,
these methods introduce a finite thickness of the membrane, which
seems, up to now, to set a quite severe limitation regarding
extraction of quantitative data in the dynamical regimes. This
requires a finite element technique with a grid refinement. Other
types of methods consist of solving the fluid equations by adopting
a "coarse-grained or mesoscopic" technique. Examples include the
so-called multiparticle collision dynamics (MPCD) or stochastic
rotation dynamics (SRD)
\cite{Malvanets1999,Noguchi2005}. Its advantages are the
relative ease of implementation and inherent thermal fluctuations
which make the method very efficient if these are required.

In this paper we apply an alternative mesoscopic method, namely the
lattice-Boltzmann (LB) method. In the spirit of the LB method, a fluid is
seen as a cluster of pseudo-fluid particles, that can collide with each
other when they spread under the influence of external applied forces.
Advantages of the LB method are its relative ease of implementation
together with  its versatile adaptability to quite arbitrary geometries.
The LB method has been already adapted and used to perform simulations of
deformable particles such as capsules \cite{Zhang2007}, vesicles and red
blood cells \cite{Dupin2007} under flow. The main issue of the work
presented in \cite{Dupin2007} is to accomplish simulations with a large
number of particles while using a small number of nodes to reduce the
computational time and the required memory. This has been achieved by
using \textit{ad hoc} membrane forces that penalize any deviation from the
equilibrium configuration.  In the present paper we use the precise
analytical expression of the local membrane force as it has been derived
\cite{Kaoui2008} from the known Helfrich bending energy
\cite{Helfrich1973}. The perimeter conservation in our case is achieved by
using a field of local Lagrangian multiplicators (equivalent to an effective
tension). To accomplish the fluid-vesicle coupling we follow the same strategy
used in \cite{Zhang2007} to simulate capsules dynamics. In \cite{Zhang2007} the
flow is computed by LB. The flow-structure two-way coupling is achieved using
the Immersed Boundary Method (IBM).
Although confining walls were considered in the above mentioned studies their
effect on the dynamics was not studied. We believe that it is of interest to
study the impact of the walls on the dynamics of vesicles, a question that to
the best of our knowledge has not been treated in the literature so far for
vesicles, capsules or red blood cells, but only for a droplet
\cite{Janssen2007} and a hard sphere \cite{Sangani2010}.

Vesicles are closed lipid membranes encapsulating a fluid and are
suspended in an aqueous solution. Their membrane is constituted of lipid
molecules (also the major component of the RBC membrane)
\cite{Lipowsky1995}. Each one has a hydrophilic head and a two hydrophobic
tails. These molecules re-organize themselves if they are in contact with
an aqueous solution, or properly speaking self-assemble, into a bilayer in
which all the heads of the molecules are facing either the internal fluid
or the external one. Experimentally, vesicles with size of the order of
$10 \mu m$ - called Giant Unilamellar Vesicles (GUV) - can be easily
prepared in the laboratory using, for example, the electro-formation
technique \cite{Angelova1992}. Unlike RBCs, for vesicles we can vary their
intrinsic characteristic parameters (size, degree of deflation, nature of
internal fluid, etc...). Despite the simplicity of their structure,
vesicles have exhibited  many features observed for red blood cells:
equilibrium shapes \cite{Seifert1991}, tank-treading motion
\cite{Fischer1978, Kraus1996}, lateral migration
\cite{Secomb2007,Kaoui2008,Coupier2008}, or slipper-like shapes
\cite{Skalak1969,Kaoui2009a}. Capsules (a model system incorporating shear
elasticity) have also revealed some common features with vesicles
\cite{Bagchi1,Bagchi2}.

In the following sections we briefly introduce the formulation of the LB
method for vesicles.  We then study in two-dimensions (2D) the
tank-treading motion of a single vesicle under shear flow between two parallel
plates. Here we decided for 2D simulations since they are computationally less
demanding, but still capture all the relevant physics. We use large systems (in
lattice units) because of the higher resolution required to extract the results
shown below. Previous works done in 2D dealing with vesicles (also for red
blood cells) have demonstrated that the dynamics in the third dimension is not
relevant even in confined geometries \cite{Secomb2007,Coupier2008, Finken2008}.
Vesicle dynamics under shear has been extensively studied
in the literature. It is known that a vesicle placed in shear flow
performs different kinds of motions depending on its degree of deflation,
the viscosity contrast between the internal and the external fluids and on
the strength of the shear flow (see the phase-diagram in
\cite{Kaoui2009b}). When the viscosities of the internal and the external
fluids are identical the vesicle performs a \textit{tank-treading} motion.
Its main long axis gets a steady inclination angle with respect to the
flow direction while its membrane undergoes a tank-treading like motion.
However, in the majority of the previous theoretical and numerical works
the vesicle is placed in an infinite fluid (unbounded domain). This
corresponds to the situation where the walls are too far from the vesicle
to have any influence on its dynamics. For this reason here we study
vesicle dynamics in a confined geometry. However, studying numerically the
dynamics of vesicles in such conditions is a challenging problem from a
computational point of view, especially in highly  confined situations. We
need to solve for the flow of the internal and the external fluids. The
boundary separating the two fluids is also an unknown quantity since the
membrane shape is not known \textit{a priori}.

Since the vesicle size ($\sim 10\mu m$) is much larger than its membrane
thickness ($\sim 5 nm$), mathematically we model the membrane as an
interface with zero thickness. Tracking the motion of this freely moving
interface under flow is not a simple task, especially when the membrane
undergoes larger deformations due to hydrodynamical stresses. We need to
label the interface by points which we track in time. Further, to take
into account the deformation an increased number of label points is
required for the code to be stable and to capture deformation with good
resolution. On the other hand, spatial derivatives on the membrane are
needed to be evaluated at every time step to compute the membrane force.
We need to evaluate the local curvature that is the fourth derivative of
the vector position. Any formation of a highly buckled region in the
membrane will introduce potential  instability. Furthermore, the vesicle
volume (the enclosed area in 2D) and its surface area (the perimeter in
2D) have to be kept conserved in time. At higher degrees of confinement
possible undesirable contact between the membrane and the walls of the
channel can be expected, and this is an additional difficulty to cope
with. We do not use any {\it ad hoc} repulsive force from the wall, rather
the non contact is achieved via a proper handling of the viscous lubrication
forces by the lattice Boltzmann method.

We shall discuss how the vesicle-fluid coupling is accomplished. For that
purpose an approach known from the immersed boundary method
\cite{Mittal2005} is adopted.  We present tests of the code by
investigating  vesicle equilibrium shapes in a fluid at rest.  We then
present simulation results regarding  the steady inclination angle and the
effective viscosity, as well as the tank-treading velocity  as  functions
of the reduced volume and the degree of confinement.
\section{The lattice-Boltzmann method}
The motion of the membrane can be induced by exerting an externally
applied flow, and this is the physical situation we are interested in.
In the present section we discuss how the fluid flow is solved for by
using the LB method. In recent decades, the LB method has been introduced
and widely used to simulate e.g. fluid flow in complex geometries (e.g. in
porous media), multi-component and multi-phase flow (e.g droplets and
binary fluids) \cite{Succi2001,Sukop2006}. Such popularity of the LB
method among scientists and engineers has been gained thanks to its
straightforward implementation and its local nature  that allows for
parallel programming.

In the limit of small Mach $Ma$ (ratio of the speed of a fluid particle in
a medium to the speed of sound in that medium) and Knudsen $Kn$ (ratio of
the molecular mean free path to the macroscopic characteristic length
scale) numbers the LB method is known to recover with good approximation the Navier-Stokes
equations \cite{Succi2001,Sukop2006}:
\begin{eqnarray}
\rho \left( \frac{\partial \textbf{u}}{\partial t}+\textbf{u}\cdot \nabla \textbf{u} \right) &=& -\nabla p + \eta \nabla^{2}\textbf{u}+\textbf{F},\\
\nabla \cdot \textbf{u} &=& 0,
\label{eq:NS_Eqs}
\end{eqnarray}
governing the fluid flow of an imcompressible Newtonian fluid. $\rho$ and
$\eta$ are the mass density and the dynamic viscosity of the studied
fluid, $\textbf{u}$ and $p$ are its velocity and pressure fields, and $t$
is the time.
$\textbf{F}$ on the right-hand side is a bulk force (e.g. gravity) or the
membrane forces as is the case for vesicles immersed in that fluid (see
below). In the spirit of the LB method, a fluid is seen as a cluster of
pseudo-fluid particles, that can collide with each other when they spread
under the influence of external applied forces. In the LB context, not
only the spatial position is discretized but also  the velocity. This
implies that every pseudo-fluid particle can move just along discrete
directions with given discrete velocities. The main quantity associated
with a pseudo-fluid particle is the distribution function
$f_{\text{i}}(\textbf{r},t)$, with $0 \leq f_{\text{i}} \leq 1$, which
gives the probability of finding at time $t$ the pseudo-fluid particle at
position $\textbf{r}$ and having  velocity $\textbf{c}_{\text{i}}$, in the
$i$-direction. There is no unique way in the choice of a lattice in the LB
method. What matters is that the discretization  has to fulfill the
following constraints: mass conservation, momentum conservation and
isotropy of the fluid. Here, we adopt the so-called the D2Q9 lattice,
where D2 is an abbreviation for  two-dimensional space while  Q9 refers to
the number of possible discrete velocity vectors \cite{Qian1992}.

The evolution in time of the distribution $f_{\text{i}}$ is governed by
the LB equation:
\begin{equation}
f_{\text{i}}(\textbf{r}+\textbf{c}_{\text{i}}\Delta t,t+\Delta t)-f_{\text{i}}(\textbf{r},t)=\Delta t \left( \Lambda _{\text{i}}+F _{\text{i}}\right) \quad (i=0...8),
\label{eq:Boltzmann_Eq}
\end{equation}
where $f_{\text{i}}(\textbf{r},t)$ is the old distribution of the
pseudo-fluid particle when it was at position $\textbf{r}$ at previous
time $t$, $f_{\text{i}}(\textbf{r}+\textbf{c}_{\text{i}}\Delta t,t+\Delta
t)$ is the new distribution of the same pseudo-fluid particle after it
moved in the direction $\textbf{c}_{\text{i}}$ to the new location
$\textbf{r}+\textbf{c}_{\text{i}}\Delta t$ during the elapsed time $\Delta
t$, with $\Delta t$ being the time step. The grid spacing is referred to
by $\Delta x$. In this paper all units are given in lattice-units, where
$\Delta x= \Delta t = 1$. The left-hand side of
Eq.~(\ref{eq:Boltzmann_Eq}) alone represents  the free propagation  of the
pseudo-fluid particles without externally applied forces. In the
right-hand side of Eq.~(\ref{eq:Boltzmann_Eq}), $F_{\text{i}}$ is any
externally applied force and $\Lambda _{\text{i}}$ is the collision
operator. Here, we adopt the Bhatnagar-Gross-Krook (BGK) approximation
which is given by
\begin{equation}
 \Lambda _{\text{i}}=-\frac{1}{\tau}\left[ f_{\text{i}}(\textbf{r},t) - f_{\text{i}}^{\text{eq}}(\textbf{r},t) \right].
\end{equation}
The BGK collision operator describes the relaxation of the distribution
$f_{\text{i}}(\textbf{r},t)$ towards an equilibrium distribution
$f_{\text{i}}^{\text{eq}}(\textbf{r},t)$, with a relaxation time $\tau$.
This relaxation time is set to 1 in this paper and related to the dynamic viscosity $\eta$ via the
relation:
\begin{equation}
\eta = \nu \rho = \rho c_{\text{s}}^{2} \frac{\Delta x^2}{\Delta t}\left( \tau - \frac{1}{2}\right),
\end{equation}
where $c_{\text{s}}=1/ \sqrt{3}$ is the speed of sound for the D2Q9
lattice. $f_{\text{i}}^{\text{eq}}(\textbf{r},t)$ is the equilibrium
distribution obtained from  an approximation of the Maxwell-Boltzmann
distribution and  is given by
\begin{equation}
f_{\text{i}}^{\text{eq}}(\textbf{r},t)\!=\!\omega_{\text{i}}\rho(\textbf{r},t)\!\left[\! 1
\!+\! \frac{1}{c_s^2}\left( \textbf{c}_{i}\! \cdot\! \textbf{u}\right)
\!+\! \frac{1}{2c_s^4}\left( \textbf{c}_{\text{i}} \!\cdot\! \textbf{u}\right)^2
\!-\! \frac{1}{c_s^2}\left( \textbf{u}\! \cdot\! \textbf{u}\right)\!
\right]\!,
\end{equation}
where $\omega_{\text{i}}$ are weight factors. $\omega_{\text{i}}$ equals
$4/9$ for the $0$ velocity vector, $1/9$ in the horizontal and vertical
directions and $1/36$ in the diagonal directions. The macroscopic
quantities describing the flow are given by
\begin{equation}
\rho(\textbf{r},t)=\sum_{\text{i}=0}^{8}f_{\text{i}}(\textbf{r},t),
\end{equation}
for the local mass density,
\begin{equation}
\textbf{u}(\textbf{r},t)=\frac{1}{\rho(\textbf{r},t)}\sum_{\text{i}=0}^{8}f_{\text{i}}(\textbf{r},t)\textbf{c}_{\text{i}},
\end{equation}
for the local fluid velocity and
\begin{equation}
p(\textbf{r},t)=\rho(\textbf{r},t)c_{\text{s}}^{2},
\end{equation}
is the local fluid pressure.

The computational domain is a rectangular box, with length $2L$ and height
$2W$. We use $x$ for the horizontal position of the box and $y$ for the
vertical position. Periodic boundary conditions are imposed on the right
and on the left side of the box. To generate a shear flow, the upper and
lower walls are displaced with the same velocity $\textbf{u}_{wall}$ but
in opposite directions. To achieve this numerically, within the LB
technique, the following bounce back boundary conditions are implemented
on the two walls as \cite{Ladd1994}
\begin{equation}
f_{-i}(\textbf{r},t+\Delta t) = f_i(\textbf{r},t) + 2\frac{\rho w_i}{c_s^2}(\textbf{u}_{wall} \cdot \textbf{c}_1).
\end{equation}
Here, $f_{-i}$ denotes the distribution function streaming in the opposite
direction of $i$. In the absence of  a vesicle, the flow relaxes towards a
steady linear shear velocity profile of the form $ \textbf{u}^\infty =
\gamma y \textbf{c}_1$, where $\gamma = u_{wall} / W$ is the shear rate.
\section{Fluid-vesicle interaction}
We denote by $\Omega_{\rm ext}$, $\Omega_{\rm int}$ the fluid domains
outside and inside the vesicle, respectively, and by $\Gamma$ the vesicle
boundary.  The flow has to be computed considering boundary conditions on
the membrane, which is a free moving interface. At the membrane $\Gamma$
we require the continuity of the flow velocity
\begin{equation}
\textbf{u}^{\text{ext}}(\textbf{r}_{\text{m}})
=\textbf{u}^{\text{int}}(\textbf{r}_{\text{m}})
=\textbf{v}(\textbf{r}_{\text{m}}), \quad \text{with} \quad \textbf{r}_{\text{m}}\in \Gamma.
\end{equation}
The $ext$ and $int$ suffixes are for the external and the internal fluids,
respectively. $\textbf{v}$ is the velocity of any point $\textbf{r}_m$
belonging to the membrane. The continuity of the tangential  velocities of
the two fluids on each side of the membrane follows from the assumption of
the no-slip boundary condition at the membrane. Continuity of the normal
velocity is a consequence of mass conservation (integrating  the
incompressibility condition  $\nabla \cdot \textbf{u} = 0$ on a small
volume straddling membrane and using the divergence theorem yields that
condition). Continuity of the two fluid velocities with that of the
membrane expresses the fact that the membrane is non permeable (normal
component) and that we assume full adherence (tangential component)
\cite{Cantat2003}. Force balance (in the absence of inertia) implies that
the net force acting on a membrane element is zero:
\begin{equation}
\left( \sigma ^{\text{ext}}(\textbf{r}_{\text{m}}) - \sigma
^{\text{int}}(\textbf{r}_{\text{m}}) \right) \textbf{n} = -
\overline{\textbf{f}}(\textbf{r}_{\text{m}}), \quad\!\! \text{with}\!\!  \quad
\textbf{r}_{\text{m}}\in \Gamma.\!
\end{equation}
$\sigma$ is the hydrodynamical stress expressed by $\sigma _{ij} = \eta
(\partial _{i} u_{j} + \partial _{j} u_{i})-p\delta _{ij}$ and
$\textbf{n}$ the unit vector normal to the membrane, pointing from the
interior domain  of the vesicle to the exterior one.
$\overline{\textbf{f}}$ is the force exerted by the membrane on its
surrounding fluid and its expression is given below. At sufficiently large
distance from the vesicle membrane, the perturbation of the velocity field
due to the membrane decays so that the fluid flow recovers its undisturbed
pattern:
\begin{equation}
\textbf{u}^{\text{ext}}(\textbf{r}) \underset{ \vert \textbf{r} - \textbf{r}_{\text{m}} \vert \rightarrow \infty}{\longrightarrow} \textbf{u}^{\infty}(\textbf{r}),
\end{equation}
where $\textbf{r}_{\text{m}}\in \Gamma$ and $\textbf{r} \in \Omega _{ext}$.

\begin{figure*}
{\includegraphics[width=0.7\textwidth]{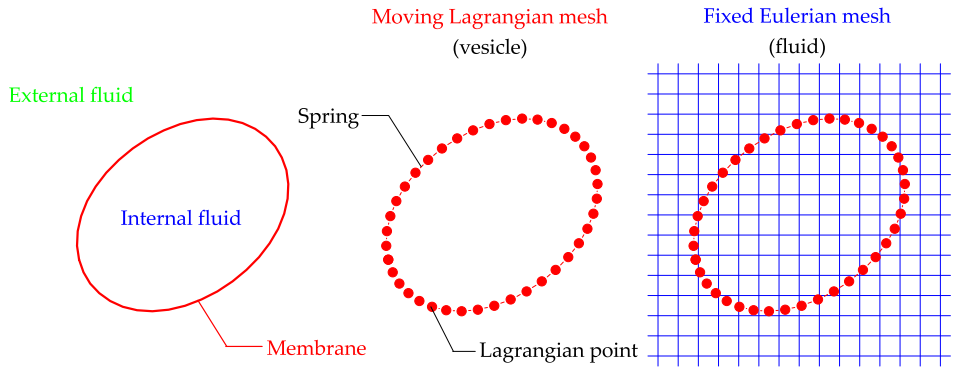}}
\caption{\label{fig:ibm1}%
(Color online). Schematic view of  a moving Lagrangian mesh representing a two-dimensional
vesicle (where the membrane is represented by a contour) immersed in a
fixed Eulerian mesh representing a fluid.}
\end{figure*}

In what follows we show how these boundary conditions can be used to
achieve the coupling between the fluid flow and the vesicle dynamics. In
the present work the internal and the external fluid flows are computed by
the LB technique. The velocity and the pressure fields are computed on an
Eulerian regular fixed mesh, while the vesicle membrane is represented by
a Lagrangian moving mesh immersed in the previous fluid mesh. The adopted
method is the so-called immersed boundary method (IBM). This method was
developed by Peskin to simulate blood flow in the heart \cite{Peskin1977}.
It is an adequate method to simulate deformable structures in flow
(fluid-structure interaction). For a review see for example Ref.
\cite{Mittal2005}. Within the framework of this method an interface
(separating two regions occupied by two distinct fluids) is discretized
into points interconnected by elastic 'springs' (as  illustrated in
Fig.~\ref{fig:ibm1} for the case of a vesicle).
First, the fluid flow is computed in the whole computational domain by
ignoring the existence of the interface. Then, the interface is advected
by the actual fluid velocity, obtained from the Eulerian mesh by
interpolation, as explained below. The fluid feels the existence of the
vesicle due to singular point forces exerted by the interface nodes on
their respective surrounding fluid nodes. This is achieved by linking the
physical quantities computed on each mesh using a so-called discrete delta
function suggested by Peskin \cite{Peskin2002}. The discrete delta
function is defined as
\begin{equation}
\Delta(\mathbf x) =
\frac{1}{16\Delta x^{2}}\left(1+\cos\frac{\pi x}{\Delta x}\right)\left(1+\cos\frac{\pi y}{2\Delta x}\right)
\label{eq:dirac}
\end{equation}
for $\vert x \vert\leqslant2 \Delta x$  and  $\vert y \vert\leqslant2
\Delta x$.
In all other regions we set $\Delta(\mathbf x) = 0$, so that the function
$\Delta$ has non zero values on a square. Here we choose $4\Delta x \times
4\Delta x$. The velocity at a given membrane node $\textbf{r}_m$ is
evaluated by interpolating the velocities at its nearest fluid nodes
$\textbf{r}_m$ using the above $\Delta$ function so that we obtain
\begin{equation}
\textbf{v}(\textbf{r}_m) = \sum _f \Delta (\textbf{r}_f - \textbf{r}_m)\textbf{u}(\textbf{r}_f).
\end{equation}
Here, $\textbf{u}(\mathbf{r}_f)$ is obtained from the LB procedure.
Deducing the velocity on the membrane nodes from the velocity of the fluid
nodes is possible since we consider that the fluid velocity is continuous
across the membrane and that the vesicle points are massless, behaving as
tracer-like particles, which do not disturb the flow at this stage. After
evaluating every membrane node velocity we update its position using an
Euler scheme
\begin{equation}
\textbf{r}_m(t+ \Delta t) = \textbf{r}_m(t) + \textbf{v}(\textbf{r}_m(t)),
\end{equation}
and consequently the vesicle is advected and deformed. However, the
vesicle membrane is not a passive interface. It reacts back on the flow
thanks to its restoring bending force
\begin{equation}
\overline{\mathbf{f}}({\mathbf r}_{m}) \!=\! \left[\! \kappa _B \!\left( \frac{\partial^2 H}{\partial s^2} +
\frac{{H}^3}{2}\right) \!-\! H \zeta \right]\!\mathbf{n} +
\frac{\partial\zeta}{\partial s}\mathbf{t} + \kappa _A\left( A \!-\!
A_0\right)\mathbf{n} ,\!
\label{eq:force}
\end{equation}
where $H$ is the local membrane curvature, $\kappa _{B}$ is the bending
modulus, $s$ is the arclength coordinate along the membrane (the contour
in 2D), $\mathbf{n}$ and $\mathbf{t}$ are the normal and the tangential
unit vectors, respectively. $\zeta$ is a Lagrange multiplier field that
enforces local length conservation (the membrane is a one dimensional
incompressible fluid). A detailed derivation of this force can be found in
\cite{Kaoui2008}. The additional last term in Eq.~(\ref{eq:force}) is
introduced in order  to enforce area conservation, because numerically a
slight variation is observed (see Ref.~\cite{Cantat2003}). $A_0$ is the
initial reference enclosed area of the vesicle, $A$ is its actual area and
$\kappa _A$ a parameter that is chosen in a such a way to keep the vesicle
area conserved. This conservation constraint can be improved also by
increasing the resolution. The membrane force has non zero value only on
the membrane and should vanish elsewhere. More precisely, a given fluid
point $\textbf{r}_f$ is subject to the force
\begin{equation}
{\bf F}({\textbf{r}_f}) = \int _{\Gamma} {\overline{\mathbf{f}}
(\textbf{r} _m)}\delta({\textbf{r} _f - \textbf{r} _m})ds(\textbf{r} _m),
\quad \text{with}  \quad \textbf{r} _m \in \Gamma , \label{eq:coupling2}
\end{equation}
where $ds$ is the distance between two adjacent membrane points. However,
since the membrane is discretized and thus presented by a cluster of
points, this integral is rather a sum of the singular forces localized on
the membrane nodes. In addition, writing the force felt by a fluid node in
terms of an ordinary Dirac delta function is not adapted here, since the
membrane nodes can be off-lattice and do not necessarily coincide with the
fluid lattice nodes. The Dirac delta function in Eq. (\ref{eq:coupling2})
is replaced by the $\Delta$ function suggested above which has a peak on
the membrane node and decays at a distance equal to twice the lattice
spacing after which it vanishes \cite{Peskin2002}. In this way the
membrane force has a non zero value in a squared area of $4 \Delta x
\times 4 \Delta x$ centered on the membrane node. The force then takes the form
\begin{equation}
{\bf F}({\textbf{r}_f}) = \sum _{m = 1}^{n} {\overline{\mathbf{f}} (\textbf{r}_m)}\Delta({\textbf{r}_f - \textbf{r}_m}).
\label{eq:coupling3}
\end{equation}
Here, $n$ is the number of  membrane nodes. The vesicle membrane finds
itself, on the one hand,  advected by its surrounding fluid and on the
other hand it exerts a force in response to the applied hydrodynamical
stresses,  causing thereby a disturbance and modification of the fluid
flow.
\section{Simulation results and discussion}
\subsection{Dimensionless numbers}
The fluid flow and vesicle dynamics are controlled by the following
dimensionless parameters:
\begin{figure*}
{\includegraphics[width=1.\textwidth]{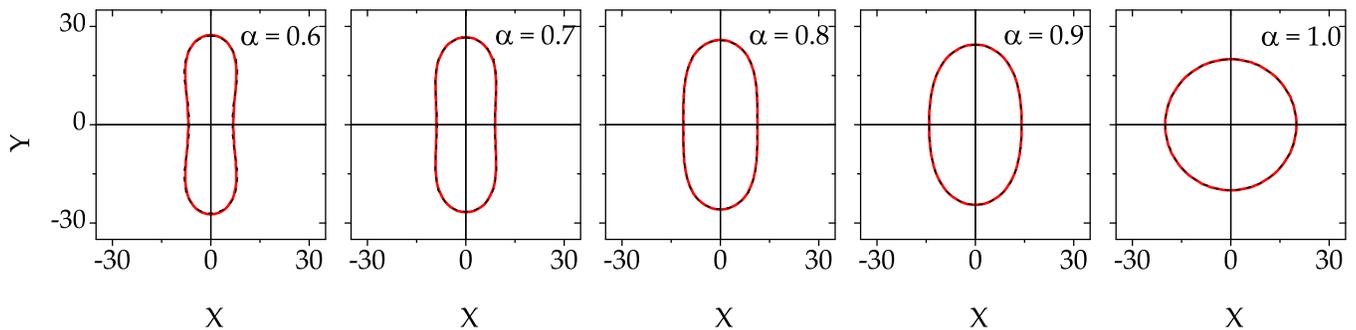}}
\caption{\label{fig:shape2}
(Color online). Computed equilibrium shapes for vesicles having the same perimeter, but
different reduced volumes $\alpha$. Red colored solid lines are shapes
computed by the LB method. For comparison purpose and for validation, we
plot also equilibrium shapes computed by the boundary integral method
\cite{Beaucourt2004} (black dashed line).}
\end{figure*}
\begin{itemize}
\item The \textit{Reynolds number}
\begin{equation}
\mbox{Re} = \frac{\rho \gamma R_0^2}{\eta},
\label{eq:Reynolds}
\end{equation}
is associated to the applied shear flow and measures the importance of the
inertial forces versus the viscous ones. $R_0$ is the effective vesicle
radius. In 2D, $R_0$ can be deduced from the vesicle perimeter $R_0 =
P/2\pi$. In our simulations we use small enough values for $\mbox{Re}$
(see below).
\item The \textit{capillary number}
\begin{equation}
Ca= \frac{\eta \gamma R_0^3}{\kappa_B},
\end{equation}
represents the ratio between the shear time ($1/\gamma$) and the
characteristic time ($\eta R_0^3 / \kappa _B$) needed by a vesicle to
relax towards its equilibrium shape after flow cessation. This parameter
controls the deformability of the vesicle under flow. Larger $Ca$
lead to a larger deformability. Below we use the value  $Ca = 1$ which
corresponds to the intermediate regime.
\item The \textit{reduced volume} (the swelling degree) quantifies how much a vesicle is swollen. In two dimensions it is given by
\begin{equation}
\alpha = \frac{A}{A_C} = \frac{4\pi A}{P^2},
\end{equation}
where $A_C$ is the area of a circle having the same perimeter $P$ as the
vesicle. $\alpha$ is unity  for a circular vesicle (a maximally swollen
vesicle) and less than unity for a deflated one $0 < \alpha < 1$.
\item The \textit{viscosity ratio} between the internal and external
fluids is given by
\begin{equation}
\lambda = {\eta_{int}\over \eta_{ext}}.
\end{equation}
In this paper, however, this ratio is taken to be unity. For this value, a
vesicle is expected to undergo tank-treading motion
\cite{Keller1982,Kraus1996,Kantsler2005}.
\item The \textit{tension number}
\begin{equation}
Ca _s = \frac{\eta \gamma R_0}{\kappa _P}
\end{equation}
is the ratio between the spring relaxation time (recall that $\kappa _P$
is the spring constant) and the shear time ($1/ \gamma$). This number
controls the inextensibility of the membrane under flow. To ensure the
vesicle perimeter conservation constraint we set $Ca_s$ significantly
small as compared to $Ca$ (below we set
$Ca_s = 1.05 \times 10^{-5}$). For the simulations, we tune $\kappa _P$
until we get very negligible variations of the perimeter $P$. $\kappa _P$
is related to $\xi$ (the Lagrange multiplier) via the formula $\xi(s,t) =
\kappa _P \left[ ds(s,t) - ds(s,0)\right]$, where $ ds(s,0)$ is the
initial reference value \cite{Cantat2003,Kaoui2008}.
\item The \textit{degree of confinement} is given by the ratio of the
vesicle's effective radius to the channel half height,
\begin{equation}
\chi = R_0/W.
\end{equation}
\end{itemize}
\subsection{Computed equilibrium shapes}
Finding the vesicle equilibrium shapes constitutes one of the benchmarking
tests we use to validate our code. In contrast to a droplet, which adopts
spontaneously a spherical equilibrium shape, vesicles can adopt different
kinds of non-spherical shapes. In two-dimensions, a vesicle gets a
circular equilibrium only for $\alpha=1$. Usually the equilibrium shapes
are obtained by minimizing the Helfrich bending energy \cite{Helfrich1973}
\begin{equation}
E=\frac{\kappa _B}{2}\int_\Gamma (2H)^2 ds,
\label{eq:Helfrich}
\end{equation}
subject to the two constraints of vesicle area $A$ and perimeter $P$
conservation (in 2D). The only parameter controlling the shape of a
vesicle, in the absence of an external applied flow and in unbounded
domain, is its reduced volume $\alpha$ \cite{Seifert1991}. An alternative
to energy minimization is to set the flow to zero ($Re=0$) and let the
vesicle relax to its terminal shape. Technically, we place initially a
vesicle with some shape (here an ellipse) in a fluid at rest (no shear
flow). Then, the membrane starts to deform in order to relax towards the
shape that minimizes its energy (Eq.~\ref{eq:Helfrich}). During this
transition the membrane induces some weak fluid flow, inside and outside
the vesicle. This flow stops once the vesicle gets its equilibrium shape.
Figure \ref{fig:shape2} shows the computed equilibrium shapes for five
vesicles having different values of the reduced volume $\alpha = 0.6$,
$0.7$, $0.8$, $0.9$, and $1$. The five vesicles have the same perimeter.
Varying the reduced volume is achieved only by varying the vesicle area.
It is somehow like swelling or deflating these vesicles to get different
equilibrium shapes. To perform simulations we used the physical parameters
$Re = Ca = 0$ (fluid at rest), $R_0 = 20$ (to achieve a sufficient
resolution at the scale of the LB grid) and $\chi = R_0 / W = 0.1$.
We have set $n = 100$, a value for which  the code is stable.
Significantly larger values of $n$ may cause instability.  From this point
of view, the LB method differs from the other conventional numerical
schemes, for instance the boundary integral or the finite
difference/element methods. In those methods, higher resolution and higher
stability is achieved by increasing (without limit) the number of
discretization points. In contrast, with the LB method an increase of $n$ induces higher
resolution, but care should be taken not to exceed some given threshold
value, beyond which the code destabilizes~\cite{krueger-varnik-raabe:2010}. 
Therefore, in all our simulations we have kept a sufficiently
small enough number of membrane nodes per lattice grid (by keeping the
distance between two adjacent membrane nodes $ds$ close to $1$).
%This instability occurs because
%for high $n$ summing the membrane force generated by membrane points
%results into larger forces on the nearby fluid nodes, and this renders the
%code unstable (larger forces induce larger speeds on the fluid nodes).
Within the LB method the velocity has to be kept small enough (in our case
we choose the limit of $0.1$) in order to have a sufficiently low
Mach number and to ascertain the limit of neglectable fluid
compressibility. 
The other parameters are chosen as follows. We have set $L = 200$, so that
the flow perturbation due to the presence of the vesicle is negligible at
the computational domain boundaries where periodic boundary conditions are
imposed. $\kappa _A = 0.01$ to fulfil a precise enough conservation of the
vesicle enclosed area (we measure a variation of the order of $0.00015\%$)
and we set $\kappa _P = 12$ to keep the perimeter conserved as well
(variation of $0.00125\%$ is measured). The obtained shape for every given
reduced volume is compared with its corresponding shape obtained by the
boundary integral method, the black dashed lines in Fig.~\ref{fig:shape2}
(the same method as used in
Refs.~\cite{Kraus1996,Cantat1999,Beaucourt2004,Biros2008}). For a given
reduced volume, the computed equilibrium shapes obtained by both numerical
methods are indistinguishable, especially at higher values of the reduced
volume. In Fig.~\ref{fig:shape2} we can see that for a reduced volume of
$0.6$, the vesicle assumes a biconcave shape, as it is typical for
healthy red blood cells.
\subsection{Tank-treading under shear flow}
In the present section, we treat the effect of confinement on the dynamics
of a tank-treading vesicle. First we study how the physical quantities,
associated to the tank-treading regime, vary with the reduced volume.
Then, for  a given reduced volume, we analyze the effect of confinement on
dynamics and rheology. We consider a single vesicle placed in a fluid
subject to a simple shear. Here, we set $R_0=30$ in order to achieve a
high enough  resolution. For $R_0=30$, our explorations led us to the
conclusion that $n=150$ is a good compromise between numerical stability
and resolution. For this value of $n$ the code is stable even at higher
degree of confinement. This also allows us to keep a sufficient number of
fluid nodes
between the wall and the membrane, a precision required in more confined
situations. The length of the simulation box is set to $L = 600 $, chosen
as to minimize perturbations by the vesicle at the edge of the simulation
box, where periodic boundary conditions are imposed. Under such conditions
and in the absence of a viscosity contrast ($\lambda=1$) a vesicle
performs a tank-treading motion \cite{Keller1982,Kraus1996,Kantsler2005}.
It deforms until reaching a steady fixed shape with its main axis assuming
a steady inclination angle with respect to the flow direction. The
membrane undergoes a tank-treading like motion and so it generates a
rotational flow of the internal enclosed fluid.
\subsection{Effect of the reduced volume}
\begin{figure}
\resizebox{\columnwidth}{!}{\includegraphics{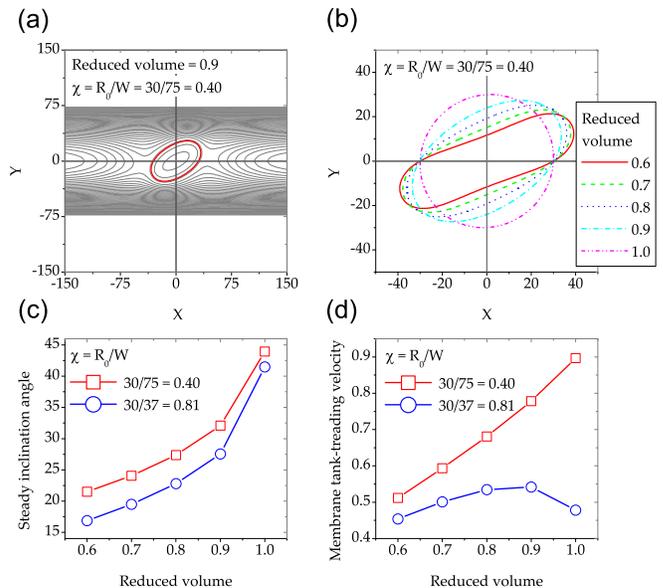}}
\caption{\label{fig:tanktreading1}%
(Color online). Physical quantities associated to the tank-treading motion of vesicles
(with $R_0 = 30$) under shear flow (with $Re = 9.45 \times 10^{-2}$ and
$Ca = 1$) in a confined channel: (a) streamlines pattern inside and outside
a vesicle ($\alpha = 0.9$) performing a tank-treading motion in a confined
channel, (b) steady shapes for different values of the reduced volumes, (c)
inclination angle versus the vesicle reduced volume for two degrees of
confinement $\chi = R_0 / W = 0.40$ and $0.81$, (d) membrane tank-treading
velocity (scaled by $\gamma R_0/2$, the rotational velocity of a circular
vesicle under shear flow in unbounded geometry) versus the reduced
volume.}
\end{figure}
\begin{figure*}
{\includegraphics[width=1.\textwidth]{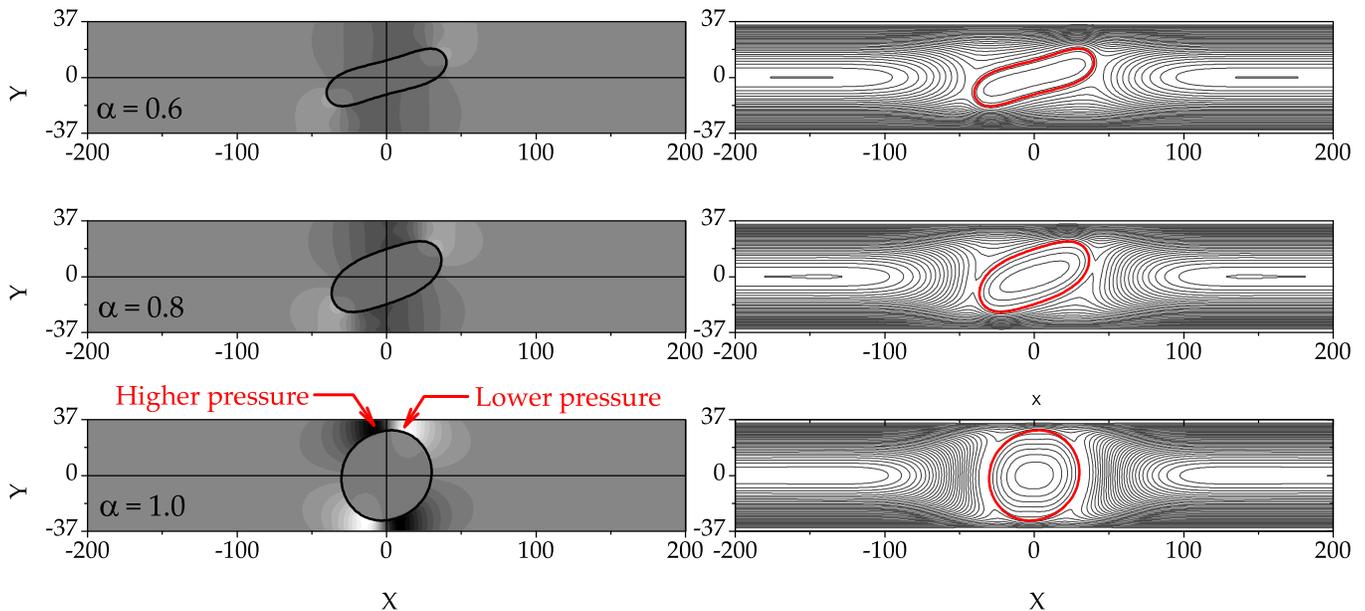}}
\caption{(Color online). Induced pressure field (with the grey scale) and flow streamlines
(the gray solid lines in the right figure), inside and outside the
vesicle, for different values of the reduced volume $\alpha = 0.6$, $0.8$
and $1$. $Re=9.45 \times 10^{-2}$, $Ca=1$ and $\chi =0.81$. The black solid
lines in the left figures and the red solid lines in the right figures
represent the vesicle membrane. In the right figures, the regions with
black color correspond to higher pressure while the white regions
correspond to lower pressure.}
\label{fig:streamlines1}
\end{figure*}
Figure~\ref{fig:tanktreading1} shows different physical quantities
measured in the tank-treading regime.  In Fig.~\ref{fig:tanktreading1}a we
show a vesicle performing tank-treading motion in a confined channel. The
vesicle assumes a steady inclination angle (the red solid line). The
streamlines inside and outside the vesicle (the gray solid lines) show
that the internal fluid undergoes   a rotational flow, induced by the
membrane tank-treading. The external fluid exhibits recirculations at the
rear (the left side of the figure) and at the front (the right side of the
figure) of the vesicle. Such recirculations do not take place  in the
unbounded geometry \cite{Biros2008}. For a tank-treading vesicle in
unbounded geometry (or at a sufficiently weak confinement), the external
fluid lines  are curved around the vesicle without being separated. In
Fig.~\ref{fig:tanktreading1}a the external fluid lines are separated into
two portions before approaching the vesicle at two saddle points (located
close to the channel centerline at the back and at the front of the
vesicle): one portion continues its flow (through the region between the
wall and the membrane) and passes the vesicle, while the other portion is
reflected back by the vesicle. Such flow recirculations are also observed
for confined rotating rigid spheres \cite{Wilson2005} and rigid ellipsoids
\cite{Aidun2000}. For the same degree of confinement ($\chi = 0.4$), in
Fig.~\ref{fig:tanktreading1}b we varied the reduced volume of the vesicle.
In Fig.~\ref{fig:tanktreading1}b we report the steady state shapes
obtained for different values of the vesicle reduced volume. All the
vesicles in Fig.~\ref{fig:tanktreading1}b have been initialized with a
zero inclination angle.  Figure~\ref{fig:tanktreading1}c shows  the steady
inclination angle as a function of the reduced volume (for two
confinements: $\chi = 0.4$ and $0.81$). The steady inclination angle
increases monotonically (for both values of $\chi $) with increasing the
reduced volume, until approaching $45$ degrees in the limiting case of
circular vesicles. The same qualitative tendency is observed in the
unbounded geometry \cite{Keller1982,Kraus1996,Beaucourt2004}.

Figure ~\ref{fig:tanktreading1}d shows the behavior of the  tank-treading
velocity normalized by $\gamma R_0/2$, which is the tank-treading velocity
of a circular vesicle \cite{Ghigliotti2009}. For $\chi =0.4$, the
tank-treading velocity increases monotonically with increasing the reduced
volume,  as observed for the unbounded geometry
\cite{Keller1982,Kraus1996,Beaucourt2004}. However, for higher
confinement, for example $\chi = 0.81$, the tank-treading velocity does
not vary anymore in a monotonous way. It has a maximum around $\alpha=
0.85$ before it decreases at larger $\alpha$. This behavior can be
explained by the fact that at higher degree of confinement, the amount of
the external fluid which is able to flow from one side (the left) to the
other side (the right) of the channel by crossing the narrow region
between the wall and the membrane becomes smaller and smaller when
increasing the reduced volume. At higher reduced volumes the inclination
angle increases and so the membrane comes in closer proximity of the wall,
see Fig.~\ref{fig:streamlines1}. Therefore, the external fluid flow does
not participate fully to generate the tank-treading motion of the vesicle.
This is also corroborated by the fact that the external fluid undergoes
recirculation (see Fig.~\ref{fig:tanktreading1}b and
\ref{fig:streamlines1}) meaning that part of the fluid is reflected
backwards when approaching the vesicle. For a  circular vesicle ($\alpha =
1$) the amount of the reflected fluid is larger compared to the amount
crossing the narrower region between the membrane and the walls.
\begin{figure}[b]
%\centerline{\includegraphics[width=0.7\columnwidth]{tanktreading4.eps}}
\resizebox{\columnwidth}{!}{\includegraphics{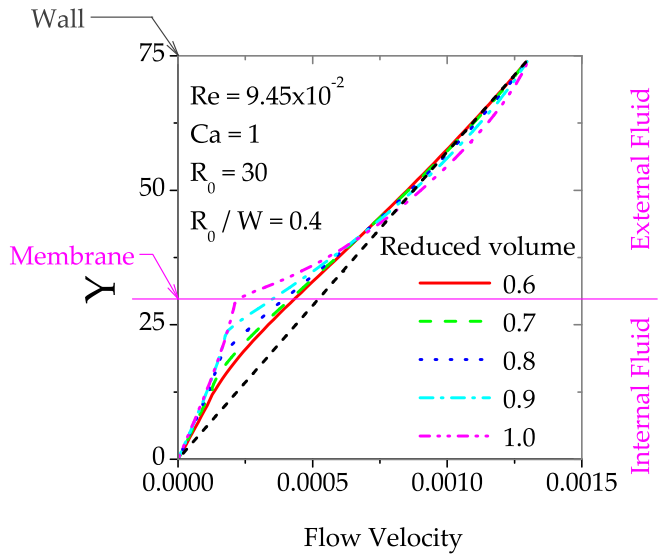}}
\caption{\label{fig:tanktreading4}
(Color online). Disturbed flow velocity profile
measured at $x=0$ for different values of the vesicle reduced volume. The
black dashed line is the undisturbed applied shear flow profile $u_x =
\gamma y$ in the absence of the vesicle. The pink solid line corresponds to the location of the membrane for $\alpha = 1$.}
\end{figure}
\begin{figure*}
{\includegraphics[width=1.\textwidth]{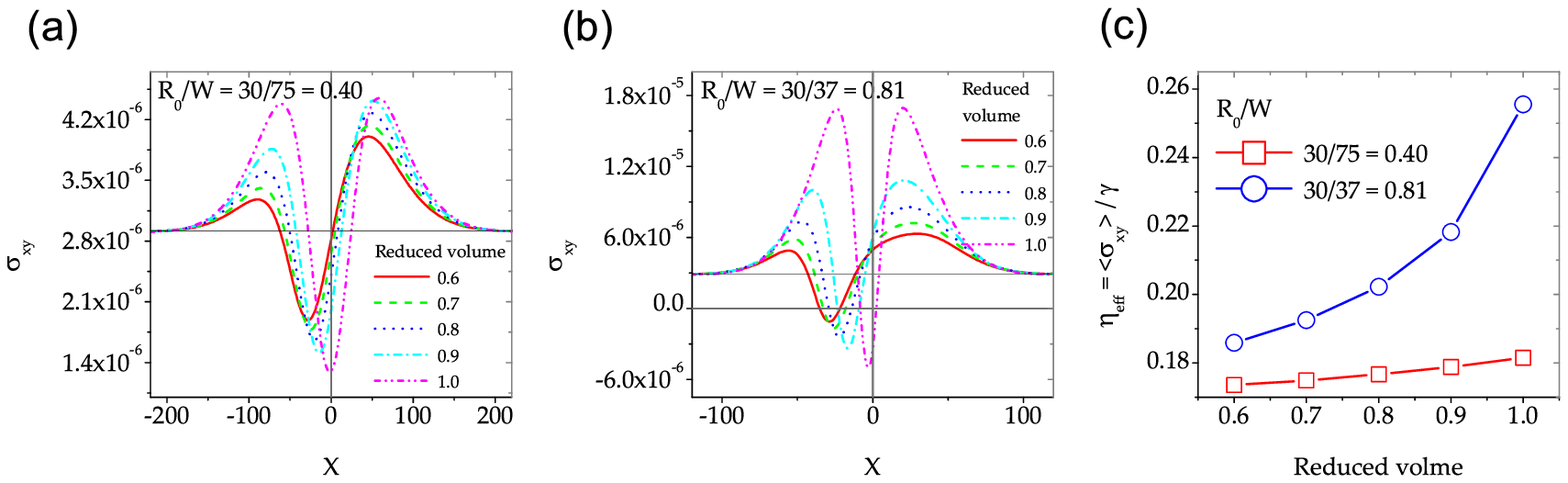}}
\caption{\label{fig:tanktreading2}
(Color online). The hydrodynamical stress exerted by the suspending fluid upon the bottom
wall for two degrees of confinement, $\chi = 0.4$ (a) and $0.81$ (b). The
grey solid line is the stress calcuted analytically in the absence of the
vesicle $\eta \gamma$. (c) The deduced effective viscosity of the fluid,
in the presence of the vesicle, for both degrees of confinement.}
\end{figure*}
The induced pressure field shown in the three left panels in
Fig.~\ref{fig:streamlines1}  is significantly  affected by increasing
$\alpha$. For $\alpha = 1$, a significant pressure gradient is observed at
the inlet and the outlet of the narrower gap between the membrane and the
wall. Such pressure drop along this narrower region generates an almost
parabolic velocity profile, for the case of $\alpha = 1$, as  is shown in
Fig.~\ref{fig:tanktreading4}. In the same figure, for comparison purpose,
we report the disturbed velocity profile for other different values of
the reduced volume (these profiles are taken at $x=0$). The disturbance is
maximal for $\alpha = 1$.

The bounce back boundary condition of Ladd \cite{Ladd1994} allows to
measure directly the hydrodynamical stress field $\sigma_{xy}$ exerted by
the fluid upon the channel walls. Figures~\ref{fig:tanktreading2}a and
\ref{fig:tanktreading2}b show the measured hydrodynamical stress for two
degrees of confinement, $\chi = 0.4$ and $0.81$. The effective viscosity
$\eta_{\rm eff}$ can be extracted from the hydrodynamical stress using the
formula
\begin{equation}
\eta_{\rm eff} = \frac{\langle \sigma_{xy} \rangle}{\gamma},
\label{eq:viscosity}
\end{equation}
where $\langle \sigma _{xy} \rangle$ refers to the volume average of the
stress tensor. The stress $\sigma _{xy}$ has been averaged along the
bottom wall. As shown in Figs.~\ref{fig:tanktreading2}a
and \ref{fig:tanktreading2}b the hydrodynamical stress on the bottom
wall exhibits important variations for a larger reduced volume
$\alpha$. For $\alpha = 1$, the stress is symmetrical with respect to the
vertical axis at $x = 0$ (which is perpendicular to the walls and passing
through the center of mass of the vesicle). This symmetry is also observed
for confined rigid spheres \cite{Sangani2010}, and is a consequence of the
symmetry of the Stokes equations upon time reversal. Actually, our
simulation contains a small amount of inertia, but so small that the
asymmetry is difficult to identify on the figure. For deflated vesicles
($\alpha \neq 1$) the stress curve has two unequal maxima and one minimum.
The flow deforms the vesicle and breaks the up-stream/down-stream symmetry
(see Fig.~\ref{fig:streamlines1} for $\alpha = 0.6$ and $\alpha = 0.8$).
In other words, the Stokes reversibility is broken by the shape
deformation. The values of these maxima and minimum significantly deviate
from the corresponding value $\eta \gamma$ in the absence of the vesicle
(presented by the horizontal grey solid line in
Fig.~\ref{fig:tanktreading2}) upon increasing the reduced volume. By
comparing Figs.~\ref{fig:tanktreading2}a and~\ref{fig:tanktreading2}b, one
notices that the stress is important for higher degrees of confinement, as
expected. Surprisingly, at higher $R_0 /W$ we observe regions with
negative hydrodynamical stress. We believe that this results from a subtle
effect due to fluid recirculation around the vesicle. However, a clear
explanation of this phenomenon is at present not available.

Figure~\ref{fig:tanktreading2}c shows the behavior of the effective
viscosity for different values of the vesicle reduced volume. The
viscosity increases when increasing the reduced volume. The same tendency
was observed for a vesicle placed in unbounded domain
\cite{Ghigliotti2009}. This result is explained as follows: for a given
confinement, the increase of the reduced volume implies a larger cross
section of the vesicle in the channel (because of the large increase in
the inclination angle), and this opposes more resistance to the fluid flow.
\subsection{Effect of confinement}
\begin{figure}
\resizebox{\columnwidth}{!}{\includegraphics{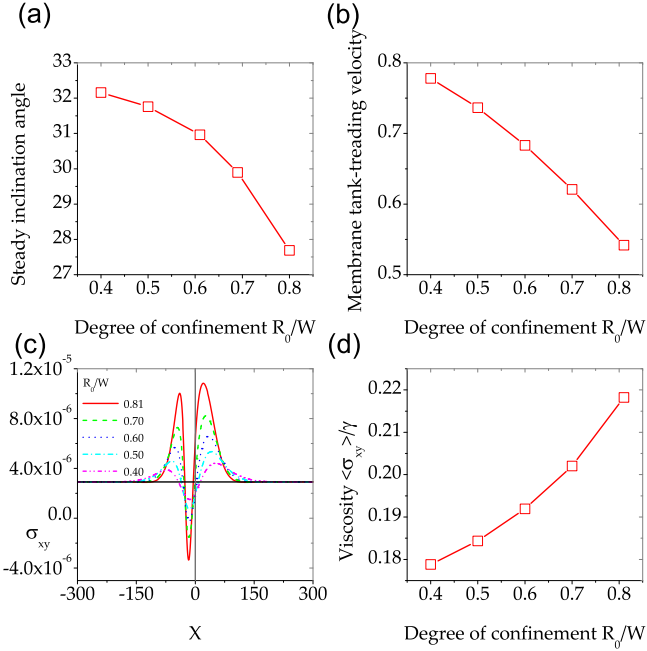}}
\caption{\label{fig:tanktreading3}
(Color online). Variation of the measured physical quantities associated to a vesicle performing tank-treading motion in confined geometries when varying the degree of confinement. (a) steady inclination angle, (b) the membrane tank-treading velocity (scaled by $\gamma R_0 / 2$), (c) the hydrodynamical stress field applied on the bottom wall and (d) the effective viscosity. Parameters are $\alpha = 0.9$, $R_0 = 30$, $Re=9.45 \times 10^{-2}$ and $Ca = 1$.}
\end{figure}
Here we set ($\alpha = 0.9$) and vary only the width ($2W$) of the
channel in order to study the effect of confinement. All the other
physical and numerical parameters are kept identical to those of the
previous section.  All simulations in Fig. \ref{fig:tanktreading3} are
performed with $\chi$ varying from $0.4$ to $0.81$. For this set of
parameters, the code is still stable and guarantees a quite satisfactory
conservation of the vesicle area and perimeter ($\Delta A / A_0 \sim
0.00015\%$ and $\Delta P/P_0 \sim 0.013\%$). Note that in Figure
\ref{fig:tanktreading3} we have kept the same shear rate $\gamma = 1.75
\times 10^{-5}$. Figure~\ref{fig:tanktreading3}a shows a decrease of the
inclination angle upon increasing  confinement. Under confinement the
angle saturates at smaller values as compared to the corresponding one in
the unbounded flow. The wall acts via a hydrodynamical repulsive force (a
viscous force) tending to push, so to speak, the orientation angle back to
zero in order to align the vesicle with the flow. Such a decrease of the
inclination angle was also reported by Janssen and Anderson
\cite{Janssen2007} for a confined droplet under shear flow.

By decreasing confinement  further, we expect to reach the unbounded
geometry limit ($\chi = 0$). We did not attempt to study this asymptotic
limit. For $\chi < 0.4$ and within the present resolution ($R_0 = 30$) a
significant increase of $L$ is required in order to avoid unphysical
effects induced by the periodic boundary conditions. This task requires a
very high amount of computational time.

\begin{figure*}
{\includegraphics[width=1.\textwidth]{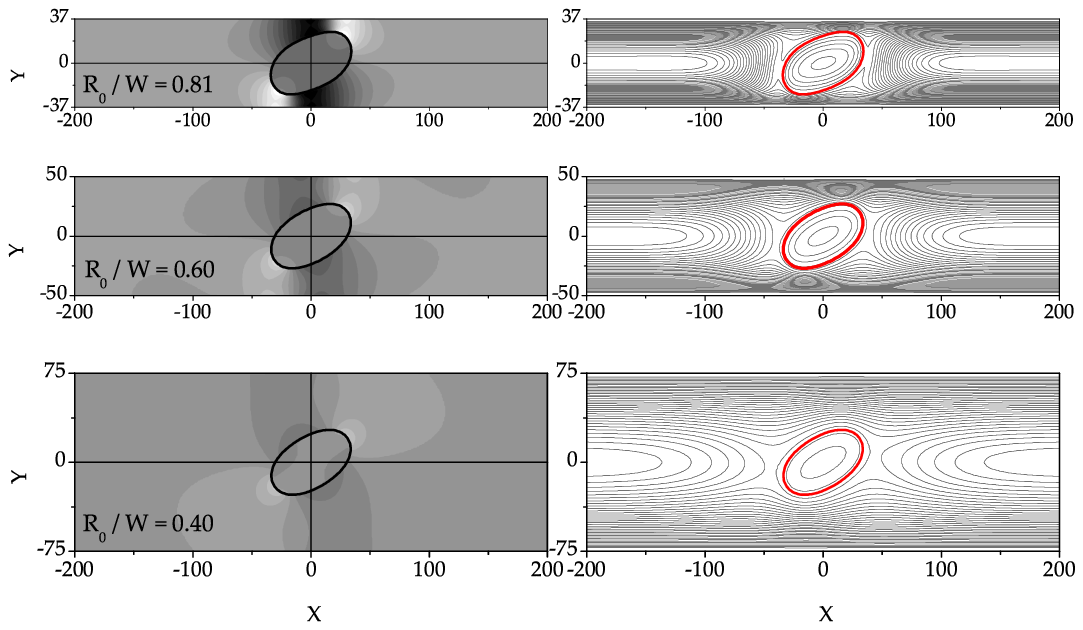}}
\caption{(Color online). Induced pressure field and flow streamlines, indside and outside
the vesicle, for different values of the degree of confinement $\chi =
0.81$, $0.6$ and $0.4$. Parameters are $\alpha = 0.9$, $R_0 = 30$,
$Re=9.45 \times 10^{-2}$ and $Ca=1$.}
\label{fig:streamlines2}
\end{figure*}

Figure~\ref{fig:tanktreading3}b  shows a decrease of the membrane
tank-treading velocity (scaled by $\gamma R_0/2$) versus confinement. At
high enough confinement, the external fluid does not anymore participate
wholly in membrane tank-treading. The fluid is partially  reflected
backwards when approaching the vesicle, an effect that increases with
confinement (see Fig.~\ref{fig:streamlines2}).
For rigid spheres a similar decrease of the rotation velocity is also
observed when increasing  confinement \cite{Sangani2010}. Varying
confinement affects also the way the membrane and the wall interact.
Figure~\ref{fig:tanktreading3}c shows the hydrodynamical stress exerted by
the fluid on the bottom wall. The horizonal black solid line is the stress
measured in the absence of the vesicle ($\eta \gamma$). At distances far
from the location of the vesicle (on the extreme right and left sides of
the figure) the stress measured for all degrees of confinement matches
with the stress in the absence of the vesicle. In the vicinity
of the vesicle, around $x=0$, we see deviation of the stress from the
value $\eta \gamma$. Such deviations become larger and larger when
increasing  confinement. Again, as in the previous section, we observe
stress with negative values at higher confinement. The effective viscosity
is extracted  from the stress (using Eq.~\ref{eq:viscosity}) and the
results are shown in Fig.~\ref{fig:tanktreading3}d. The effective
viscosity is found to significantly increase non-linearly with
confinement.

In order to gain further insight we represent the pressure field and
the streamlines (Fig.~\ref{fig:streamlines2}), for three different
confinements, $\chi = 0.81$, $0.6$ and $0.4$.  Confining further the
vesicle results in an increase of the pressure inside the vesicle,
entailing a higher pressure gradient along the fluid layer located
between the membrane and the wall. The amount of the fluid crossing
this region decreases also when increasing confinement. The
streamlines pattern shows that the recirculation becomes important
at higher degree of confinement. Their two focal  points move closer
and closer to the vesicle when increasing confinement. A closer
inspection of the pressure field and the streamlines (for a given
degree of confinement $\chi$) reveals interesting dynamics occurring
in the narrow region between the vesicle and the wall. When the
external fluid approaches the vesicle it splits into two parts. One
part is reflected by the vesicle and pushed backward without passing
the vesicle. The other part continues its flow and crosses the
narrower gap formed between the wall and the membrane. At the inlet,
of this gap, the pressure is significantly  higher, resulting in a
slowing down of the fluid (the streamlines are separated). Once the
fluid enters this region its velocity is amplified (the streamlines
come closer), under the action of the pressure gradient along the
gap, until it exits that region. At the outlet, the pressure drops
to a lower value and the fluid is slowed down again (the streamlines
separate again).

Finally, some general comments are worth mentioning. The fluid
motion in the narrower gap, formed between the vesicle membrane and
the wall, is induced under the action of three mechanisms: 1 - the
membrane force, 2 - the shear flow and 3 - the pressure-gradient
along this region. The third mechanism dominates at high
confinement. In that regime the fluid (in the gap) is subject to the
sum of forces induced by the above three mechanisms. Within the
present method, this sum must not exceed some given threshold
otherwise the code becomes unstable (due to higher flow velocities).
There is also another technical detail that becomes problematic in
situations of higher degrees of confinement. We assumed that the
membrane has a zero thickness. However, by using the expression
Eq.~\ref{eq:dirac} the membrane force is distributed on fluid nodes
located at distances of roughly $4\Delta x$ from the membrane. The
membrane acquires a non-zero effective thickness. In more confined
situations we need to leave at least $4$ fluid nodes in the gap
between the membrane and the wall, otherwise the dynamics of the
vesicle suffer from numerical artifacts. For example, a leakage of
the internal fluid is observed, and  the tank-treading velocity
exhibited a non uniform behavior along the membrane. To overcome all
these problems we had to increase the resolution. For the resolution
of $R_0 = 30$ used in this section, the upper limit of the
confinement we were able to reproduce without any apparent problem
is $\chi = 0.81$.
\section{Conclusion}
We have studied the effect of confinement between two parallel walls on
vesicle dynamics under shear flow. We limited ourselves to the case of
having the same fluid inside and outside the vesicle. In such a situation
the vesicle performs tank-treading motion. We developed a
lattice-Boltzmann method to perform two-dimensional simulations. The
coupling between fluid flow and vesicle dynamics was adopted from
the immersed boundary method. Unlike previous works, we have
introduced the membrane force by using its analytical expression as a
function of the mean curvature and its derivative. The vesicle enclosed
area and its perimeter are kept conserved in our method. We first computed
the known vesicle equilibrium shapes for different values of the swelling
degree in order  to validate our code. The obtained shapes match perfectly
the ones computed by the boundary integral method. As a second step, we
studied the case of a vesicle placed in a domain bounded by two parallel
walls. We induced the shear flow by moving these two walls in opposite
directions. We found that both the vesicle inclination angle, with respect
to the flow, and its membrane tank-treading velocity decrease when
increasing the degree of confinement. Moreover, since at sufficiently
large degree of confinement the vesicle membrane comes close to the wall
so that just a very narrow region is left for the external fluid to flow.
Therefore, the vesicle acts as an obstacle and thus the effective
viscosity increases dramatically when increasing confinement.  At a given
degree of confinement, we varied the swelling degree. We observed the same
qualitative tendency as for the unbounded geometry for the behavior of the
angle, tank-treading velocity and viscosity as a function of the swelling
degree. However at higher degree of confinement even the angle still shows
an increase with increasing the swelling degree, the measured values are
lower. The tank-treading velocity does not increase monotonically  with
the swelling degree. It exhibits a maximum value before getting to lower
values in the limit of circular vesicles. 
\section*{Acknowledgments}
We would like to thank the CNES (Centre National d'Etudes Spatiales), the
collaboration research center (SFB 716), the EGIDE PAI Volubilis (Grant
No. MA/06/144), the CNRST (Grant No. b4/015), the TU Eindhoven High
Potential Research program, and NWO/STW (VIDI grant of J. Harting) for financial support.
%
% References

%

\begin{thebibliography}{10}
%
\bibitem{Fung} Y. C. Fung, {\it Biomechanics} (Springer, New York, 1990).
%
\bibitem{Pozrikidis} C. Pozrikidis, {\it Boundary integral and singularity methods for
linearized viscous flow}, Cambridge University Press (1992).
%
\bibitem{Kraus1996}
M. Kraus, W. Wintz, U. Seifert, and R. Lipowsky,
Phys. Rev. Lett. {\bf 77}, 3685 (1996).
%
\bibitem{Cantat1999}
I. Cantat and C. Misbah,
Phys. Rev. Lett. {\bf 83}, 235 (1999).
%
\bibitem{Biros2008}
S. K. Veerapaneni, D. Gueyffier, D. Zorin, and G. Biros,
J. Comp. Phys. {\bf 228}(7), 2334 (2009).
%
\bibitem{Biben2010}
T. Biben, A. Farutin, and C. Misbah,
arXiv:0912.4702v1.
%
\bibitem{Biben2003}
T. Biben and C. Misbah,
Phys. Rev. E {\bf 67} (2003).
%
\bibitem{Biben2005}
T. Biben, K. Kassner, and C. Misbah,
Phys. Rev. E {\bf 72} 041921  (2005);
%
\bibitem{Du2006}
Q. Du, C. Liu, and X. Wang,
J. Comp. Phys. {\bf 212}, 757 (2006).
%
\bibitem{Maitre2010}
E. Maitre, C. Misbah, P. Peyla, and A. Raoult,
preprint (2010).
%
\bibitem{Malvanets1999}
A. Malevanets and R. Kapral,
J. Chem. Phys. {\bf 110}, 8605 (1999).
%
%\bibitem{Noguchi2004}
%H. Noguchi and G. Gompper,
%Phys. Rev. Lett. {\bf 93}, 258102 (2004)
%
\bibitem{Noguchi2005}
H. Noguchi and G. Gompper,
PNAS {\bf 102}(40), 14159 (2005)
%
\bibitem{Finken2008}
R. Finken, A. Lamura, U. Seifert, and G. Gompper, 
Two-dimensional fluctuating vesicles in linear shear flow. 
Eur. Phys. J. E {\bf 25}, 309 (2008)

%
\bibitem{Zhang2007}
J. Zhang, P. C. Johnson and A. S. Popel,
Phys. Biol. {\bf 4}, 285 (2007).
%
\bibitem{Dupin2007}
M.M. Dupin, I. Halliday, C.M. Care, and L.L. Munn,
Phys. Rev. E {\bf 75} 066707 (2007).
%
%\bibitem{Janoschek2010}
%F. Janoschek, F. Toschi, and J. Harting,
%Phys. Rev. E {\bf 82}, 056710 (2010).
%
\bibitem{Kaoui2008}
B. Kaoui, G.H. Ristow, I. Cantat, C. Misbah, and W. Zimmermann,
Phys. Rev. E {\bf 77}, 021903 (2008).
%
\bibitem{Helfrich1973}
W. Helfrich, Z. Naturforsch.
A \textbf{28c}, 693 (1973).
%
%\bibitem{Nash2008}
%R. W. Nash, R. Adhikari and M. E. Cates
%Phys. Rev. E {\bf 77}, 026709 (2008)
%
\bibitem{Lipowsky1995}
R. Lipowsky and E. Sackmann,
\textit{Structure and Dynamics of Membranes, from Cells to Vesicles}
(North-Holland, Amsterdam, 1995)
%
\bibitem{Angelova1992}
M. I. Angelova, S. Sol\'{e}au, P. M\'{e}l\'{e}ard, J.-F. Faucon, P. Bothorel,
Prog. Colloid Polym. Sci. {\bf 89}, 127 (1992)
%
\bibitem{Seifert1991}
U. Seifert, K. Berndl, and R. Lipowsky,
Phy. Rev. A {\bf 44}, 1182 (1991)
%
\bibitem{Fischer1978}
T. M. Fischer, M. Stohr-Liesen, and H. Schmid-Schonbein,
Science  {\bf 24}, 894 (1978)
%
\bibitem{Coupier2008}
G. Coupier, B. Kaoui, T. Podgorski and C. Misbah,
Phys. Fluids {\bf 20}, 111702 (2008)
%
\bibitem{Secomb2007}
T. W. Secomb, B. Styp-Rekowska, and A. R. Pries,
Ann. Biomed. Eng. {\bf 35},755 (2007)
%
\bibitem{Skalak1969}
R. Skalak and P. I. Branemark,
Science {\bf 164} 717 (1969).
%
\bibitem{Kaoui2009a}
B. Kaoui, G. Biros, and C. Misbah,
Phys. Rev. Lett. {\bf 103} 188101 (2009)
\bibitem{Bagchi1} S.K. Doddi, and P. Bagchi. Inter. J. Multiphase Flow {\bf 36}, 966 (2008).
\bibitem{Bagchi2} P. Bagchi and R. M. Kalluri, Phys. Rev. E {\bf 80}, 016307 (2009).
%
\bibitem{Kaoui2009b}
B. Kaoui, A. Farutin, and C. Misbah,
Phys. Rev. E {\bf 80}, 061905 (2009)
%
\bibitem{Mittal2005}
R. Mittal and G. Iaccarino,
Annu. Rev. Fluid Mech. {\bf 37}, 239 (2005)
%
\bibitem{Succi2001}
S. Succi,
\textit{The Lattice Boltzmann equation}
(Oxford University Press, 2001)
%
\bibitem{Sukop2006}
M.C. Sukop, D.T. Thorne,
\textit{Lattice Boltzmann modeling: an introduction for geoscientists and engineers}
(Springer, 2006)
%
\bibitem{Qian1992}
Y. H. Qian, D. d'Humi\`{e}res, and P. Lallemand,
Europhys. Lett. {\bf 17}, 479 (1992)
%
\bibitem{Ladd1994}
A. J. C. Ladd,
J. Fluid Mech. {\bf 271}, 285 (1994)
%
\bibitem{Cantat2003}
I. Cantat, K. Kassner, and C. Misbah,
Eur. Phys. J. E 10, 175189 (2003)
%
\bibitem{Peskin1977}
C. S. Peskin,
J. Comput. Phys. {\bf 25}, 220 (1977)
%
\bibitem{Peskin2002}
C. S. Peskin,
Acta Numerica {\bf 11}, 479 (2002)
%
\bibitem{Keller1982}
S. R. Keller and R. Skalak,
J. Fluid Mech. {\bf 120} (1982)
%
\bibitem{Kantsler2005}
V. Kantsler and V. Steinberg,
Phys. Rev. Lett. {\bf 95}, 258101 (2005)
%
\bibitem{krueger-varnik-raabe:2010}
T. Kr\"uger, F. Varnik, and D. Raabe,
Comp. Math. Appl., in press (2010)
%
\bibitem{Beaucourt2004}
J. Beaucourt, F. Rioual, T. S\'{e}on, T. Biben, and C. Misbah,
Phys. Rev. E \textbf{69}, 011906 (2004)
%
\bibitem{Wilson2005}
M. C. T. Wilson and P. H. Gaskell, and M. D. Savage,
Phys. Fluids {\bf 17}, 093601 (2005)
%
\bibitem{Aidun2000}
E-J. Ding and C. K. Aidun,
J. Fluid Mech. {\bf 423} (2000) 317
%
\bibitem{Sangani2010}
A. S. Sangani, A. Acrivos, L. Jibuti, and P. Peyla,
(unpublished)
%
\bibitem{Ghigliotti2009}
G. Ghigliotti, H. Selmi, B. Kaoui, G. Biros, and C. Misbah,
ESAIM: PROC {\bf 28}, 211 (2009)
%
\bibitem{Janssen2007}
P. J. A. Janssen and P. D. Anderson,
Phys. Fluids {\bf 19}, 043602 (2007)
%
\end{thebibliography}
\end{document}